# Modelling Concentrated Polymer Self Diffusion: An Alternative Model to Reptation

Dave E. Dunstan

*Department of Chemical Engineering, University of Melbourne, VIC 3010, Australia.*

davided@unimelb.edu.au

Herein an alternative model to reptation to describe concentrated polymer dynamics is developed. The model assumes that the chains act as blobs that are able to diffuse past each other in a compressed state. De Gennes[1] and later Doi and Edwards[2] developed the reptation model to explain observed polymer dynamics at high concentrations. The essential physics embodied in the reptation model is that chain diffusion occurs by each chain moving in a snake like motion along the contour length. The motivation for the development of the reptation model is the inability to reconcile the observed polymer diffusion with the measured macroscopic viscosities.[3] As is shown below, a physically consistent model may be developed by assuming that the local viscosity, micro-viscosity, sensed by the polymer is less than the macroscopic viscosity.

Recent experimental evidence has shown that chain compression occurs in simple Couette flow.[4,5] This is physically consistent with the observed visco-elasticity and temperature dependence of the viscosity of polymer solutions.[5] Further evidence of polymer compression in flow was found by Frith et al. using rheological measurements on sterically stabilised suspensions.[6,7]

Diffusing probe measurements on “entangled” solutions and the observed rheological behaviour, namely that the solutions flow, suggest that the reptation model is not a complete description of polymer dynamics.[3,8-10] Cai et al. interpret the diffusion of nanoparticle probes in concentrated polymer solutions as occurring via a hopping diffusion mechanism.[11] They introduce the concept of an effective activation energy for the release of the particles from the polymer cage. Nath et al. interpret several regimes that depend on the ratio of the nanoparticle to polymer radius. In solutions where the particle diameter is less than the polymer radius, the particles appear to diffuse with a reduced local medium viscosity (less than the continuum macroscopic viscosity).[12] In a further study using tracer latex particles in polymer solutions, a rapid diffusion, greater than that in the solvent, was measured.[13] Analysis of the measured fast diffusion coefficient for the tracer latex particles in polyacrylamide solutions yields an exponent of -2.25 for the particle diffusion coefficient/molecular weight relationship. The exponent is very close to that measured for the self diffusion of the polymers found for a range of systems.[14] The probe diffusion measurements are consistent with the model of the polymers as "blobs" diffusing through the surrounding solution of blobs.

The experimental evidence from the range of studies above appears to be physically inconsistent with the ideas embodied in reptation and entanglement.[1,2] If the chains are entangled in a truly ideal mixture, then it is difficult to understand how chain compression and tumbling in flow could arise. In solutions where the mesh size is less than the particle radius, "cage jumping" has been invoked to model the observed behaviour.[15] These observations bring into question the universal applicability of the reptation model.[1,2]

There exists a body of data, reviewed succinctly by Lodge[14], that shows that the self diffusion coefficients of polymers decrease with a fitted power law exponent of -2.3. This experimental evidence has been used to validate the key ideas behind reptation. The power law exponent of -2.0 has been predicted by a straightforward argument regarding the diffusion of segments of the chain along a tube that represents the contour of the chain.[3] This has been modified using the ideas of constraint release and contour length fluctuations in order to fit the value of -2.3.[14] The ideas embodied in reptation and entanglement have been considered to be the only plausible explanation of the observed diffusion, relaxation time and viscosity dependence on the molecular weight.[3] The pioneering study by Klein also measures polymer diffusion using infrared microdensitometry to measure deuterated chains diffusing into a non-deuterated melt.[16] The diffusion is observed to follow a square of the molecular weight behaviour in accord with the theory of de Gennes.[16] To date however, there exists no direct evidence that reptation is the mechanism by which polymer chains diffuse at high concentrations.

Here an alternative model to reptation for polymer self-diffusion is presented that involves the diffusion of blobs past each other. The rate at which this occurs is determined by fluctuations in the polymer blob size due to thermal fluctuations compressing the chains in an elastic manner. A key assumption in the model is that the "local" viscosity experienced by a blob is less than that of the macroscopic viscosity.

The system is defined as a concentrated solution of polymer chains in a solvent at thermal equilibrium at temperature T. For one blob to diffuse past the near neighbours, the neighbours elastically compress to enable the blob in question to escape the "cage" generated by the near neighbours. Compression of two blobs in the restricting cage to release the entrapped blob requires that each blob compresses by order $R_0/2$. The activation energy for the release of the blob from the surrounding blobs is then that determined by the compression of a chain to $R_0/2$.

Using the ideal model of the chains at high concentration, the elastic deformation energy is then:

$$F_{el} = \frac{3k_BTR^2}{2R_0^2} = \frac{9k_BT}{8} \qquad [1]$$

for a change in radius of $R_0/2$. This value of $F_{el}$ is then the energy barrier or effective activation energy, $E_a$, associated with the diffusion of the blobs.

The effective diffusion coefficient may then be written;

$$D = \frac{R_0^2}{t} \qquad [2]$$

where the time is that taken for the blob to diffuse by $R_0$.

The Arrhenius first order rate for the process is then:

$$k = \frac{D}{R_0^2} \exp\left(\frac{E_a}{k_BT}\right) \qquad [3]$$

Substitution for D and $F_{el}$ yields:

$$k = \frac{k_BT}{6\pi\eta R_0^3} \exp\left(9/8\right) \qquad [4]$$

Allowing that the viscosity η, has the form:

$$\eta \sim \phi^m \qquad [5]$$

with ϕ the volume fraction and m the power law index.

So that;

$$\eta \sim R_0^{3m} \qquad [6]$$

Then from [2], [4] and [6] above;

$$D \sim R_0^2 / R_0^{3(m+1)} \qquad [7]$$

For the case where m = 4/3 and using the ideal chain result;[17]

$$R_0 \sim N^{1/2} \qquad [8]$$

yields the form of the diffusion coefficient as:

$$D \sim N^{-5/2} \qquad [9]$$

The power law exponent of -5/2 is a reasonable approximation of the experimentally measured value of -2.3.

The value of m in Equation [5] may be used as a fit parameter for the experimentally measured exponent of -2.3 to yield a value of m = 11/9.

The viscosity and diffusion coefficient data may be fitted using reasonable values of the viscosity volume fraction relationship and reptation is not required. A problem that has been identified with this approach is that the predicted viscosity-molecular weight relationship using this model is not predicted.[3,14] The measured relationships is;

$$\eta_0 \sim M^{3.4} \qquad [10]$$

While the predicted relationship from the model above is;

$$\eta_D \sim M^2 \qquad [11]$$

Here $\eta_0$ is the measured value of the macroscopic viscosity and $\eta_D$ is the effective "diffusional viscosity". The diffusional viscosity is then that which is experienced by the individual molecules in their diffusion. If each blob is surrounded by a cage of blobs, then the relaxation time is determined by the fluctuations in the local blob radius. Each blob is assumed to compress to allow the diffusion of the blobs past each other. The local viscosity will then be a reduced contribution from the polymers plus the solvent contribution. In this case, the local viscosity will be less than the measured macroscopic viscosity and should lie between the macroscopic value of the polymer solution and the solvent. Therefore, the power law predicted by the model being less than the macroscopically measured value is physically reasonable. The square of the molecular weight dependence of the local viscosity is also reasonable, suggesting that two chains must compress in order to create an escape route for the central chain. It is worth noting that the activation energy for the release of the chains (compression by $\sim R_0/2$) is $9k_BT/8$ being very close to $k_BT$, the thermal energy in the system. As such, the activation barrier to diffusion is relatively small.

There are subtleties in that the chain may exist as a many blob chain. This changes the overall dimensions of the diffusing blobs but not the overall molecular weight dependence of the key variables. Knots and entanglements will not significantly effect the local diffusion in this sense, however knots and entanglements may play a role in the macroscopic rheological behaviour. A test of the model presented would involve measurements of probe diffusion using particles of similar radius to that of the polymers.

References.